\documentclass[prd,showpacs]{revtex4}
\begin{document}

\newcommand{\re}{\mathop{\mathrm{Re}}}

\newcommand{\be}{\begin{equation}}
\newcommand{\ee}{\end{equation}}
\newcommand{\bea}{\begin{eqnarray}}
\newcommand{\eea}{\end{eqnarray}}

\title{Do gravitational waves carry energy-momentum? A reappraisal}

\author{Janusz Garecki}
\email{garecki@wmf.univ.szczecin.pl}
\affiliation{\it Institute of Mathematics University of Szczecin
and Cosmology Group University of Szczecin,
 Wielkopolska 15, 70-451 Szczecin, Poland}
\date{\today}
\input epsf
\begin{abstract}
After direct detection gravitational radiation in 2015 many authors are publishing remakes
of their old articles about this radiation. I decided to follow this line in my Lecture
delivered at the Conference ``Varcosmofun'16'' (12-17 September
2016, Szczecin, Poland, EU). Namely, I have presented at this
Conference an updated summary of my past articles on gravitational
radiation. As a base for my presentation I have used mainly the article
published in 2002 in Annalen der Physik \cite{Gar3} and the articles
\cite{Gar4}.

In these past articles I have showed that the real gravitational waves which possess a
non-vanishing Riemann tensor always carry energy-momentum (and
also angular momentum). Our proof have used  canonical
superenergy and supermomentum tensor for gravitational field in
former articles  and  the averaged relative energy-momentum tensor
in latter. In this article we confine to the energy-momentum only.
\end{abstract}
\maketitle
\section{Introduction}
In General Relativity ({\bf GR}) the gravitational field $\Gamma^i_{~kl}$ does not possess
any energy-momentum tensor. Instead, it only possesses the so-called ``energy-momentum
pseudotensors''. In fact, this is a consequence of the Einstein Equivalence Principle ({\bf EEP})
. Because of that, many authors put in doubt the reality of the energy-momentum
(and the angular momentum also) transfer by gravitational waves. As the main argument,
some of these authors used the fact that for the majority of exact solutions of the vacuum
Einstein field equations which represent gravitational waves, energy-momentum pseodotensors
{\it globally vanish} in certain coordinates. In consequence, these pseudotensors give
``no gravitational energy and no gravitational energy flux'' in these coordinates.
Some other authors argue that the vanishing of the components $_g t^{ok}$ (or $_g
t_o^{~k}$) of the gravitational pseudotensor $_g t^{ik}$ (or $_g t_i^{~k}$) may be treated
as a coordinate condition coupled to the Einstein equations and yield (in special coordinates)
``global vanishing of the pure gravitational energy and the pure gravitational energy flux''.

However, {\it such conclusions are physically incorrect} because they relay on
non-tensorial, {\it coordinate dependent expressions} (See \cite{Gar3} for full argumentation).

The energy and the energy flux (as well as the angular momentum) of the real
gravitational field which has $R_{iklm}\not= 0$ {\it always exist and do not vanish}. In order
to show this, one should use the coordinate independent expressions like our
{\it canonical superenergy tensor} for gravitational field (used in our older articles) or
{\it canonical averaged relative energy-momentum tensor} for gravitational field
(used in our latter articles).

\section{The canonical gravitational superenergy tensor and the canonical averaged relative
gravitational energy-momentum tensor}
As it was already mentioned, in the framework of general relativity ({\bf GR}) the
gravitational field {\it has non-tensorial strengths} $\Gamma^i_{kl}
 = \{^i_{kl}\}$ and {\it admits no energy-momentum tensor}. One
 can only attribute to this field {\it gravitational
 energy-momentum pseudotensors}. The leading object of such a kind
 is the {\it canonical gravitational energy-momentum pseodotensor}
 $_E t_i^{~k}$ proposed already in past by Einstein. This
 pseudotensor is a part of the {\it canonical energy-momentum
 complex} $_E K_i^{~k}$ in {\bf GR}.

The canonical complex $_E K_i^{~k}$ can be easily obtained by
rewriting Einstein equations to the superpotential form
\begin{equation}
_E K_i^{~k} := \sqrt{\vert g\vert}\bigl( T_i^{~k} + _E
t_i^{~k}\bigr) = _F U_i^{~[kl]}{}_{,l}
\end{equation}
where $T^{ik} = T^{ki}$ is the symmetric energy-momentum tensor for matter, $g = det[g_{ik}]$,
 and

\begin{eqnarray}
_E t_i^{~k}& =& {c^4\over 16\pi G} \bigl\{\delta_i^k
g^{ms}\bigl(\Gamma^l_{mr}\Gamma^r_{sl} -
\Gamma^r_{ms}\Gamma^l_{rl}\bigr)\nonumber\cr
&+& g^{ms}_{~~,i}\bigl[\Gamma^k_{ms} - {1\over 2}
\bigl(\Gamma^k_{tp}g^{tp} -
\Gamma^l_{tl}g^{kt}\bigr)g_{ms}\nonumber\cr
&-& {1\over 2}\bigl(\delta^k_s \Gamma^l_{ml} +
\delta^k_m \Gamma^l_{sl}\bigr)\bigr]\bigr\};
\end{eqnarray}
\begin{equation}
_F {U_i^{~[kl]}} = {c^4\over 16\pi G}g_{ia}({\sqrt{\vert
g\vert}})^{(-1)}\bigl[\bigl(-g\bigr)\bigl(g^{ka} g^{lb} - g^{la}
g^{kb}\bigr)\bigr]_{,b}.
\end{equation}
$_E t_i^{~k}$ are components of the canonical energy-momentum
pseudotensor for gravitational field $\Gamma ^i_{kl} =
\bigl\{^i_{kl}\bigr\}$, and $_F {U_i^{~[kl]}}$ are von Freud
superpotentials.
\begin{equation}
_E K_i^{~k} = \sqrt{\vert g\vert}\bigl(T_i^{~k} + _E
t_i^{~k}\bigr)
\end{equation}
are components of the {\it Einstein canonical energy-momentu complex,
for matter and gravity}, in {\bf GR}.

In consequence of (1) the complex $_E K_i^{~k}$satisfies local
conservation laws
\begin{equation}
{_E K_i^{~k}}_{,k}\equiv 0.
\end{equation}
In very special cases and in special coordinates, one can obtain from these local conservation
laws the reasonable integral conservation laws of the energy and
momentum.

Despite that one can easily introduce in {\bf GR} {\it the
canonical superenergy tensor} for gravitational
field. This was done in past in a series of our articles (See,
e.g.,\cite{Gar1} and references therein).
It appeared that the idea of the superenergy tensors is universal:
to any physical field having an energy-momentum tensor or
pseudotensor one can attribute the coresponding superenergy
tensor.

So, let us give a short reminder of the general, constructive
definition of the superenergy tensor $S_a^{~b}$ applicable to
gravitational field and to any matter field. The definition uses
{\it locally Minkowskian structure} of the spacetime in {\bf GR}
and, therefore, it fails in a spacetime with torsion, e.g., in Riemann-Cartan
spacetime.
In the normal Riemann coordinates {\bf NRC(P)} we define
(pointwiese)
\begin{equation}
S_{(a)}^{~~~(b)}(P) = S_a^{~b} :=(-) \displaystyle\lim_{\Omega\to
P}{\int\limits_{\Omega}\biggl[T_{(a)}^{~~~(b)}(y) - T_{(a)}^{
~~~(b)} (P)\biggr]d\Omega\over 1/2\int\limits_{\Omega}\sigma(P;y)
d\Omega},
\end{equation}
where
\begin{eqnarray}
T_{(a)}^{~~~(b)}(y) &:=& T_i^{~k}(y)e^i_{~(a)}(y)
e_k^{~(b)}(y),\nonumber\cr
T_{(a)}^{~~~(b)}(P)&:=& T_i^{~k}(P) e^i_{~(a)}(P)e_k^{~(b)}(P) =
T_a^{~b}(P)
\end{eqnarray}
are {\it physical or tetrad components} of the pseudotensor or
tensor field which describes an energy-momentum distribution, and $\bigl\{y^i\bigr\}$
are normal coordinates. $e^i_{~(a)}(y), e_k^{~(b)} (y)$ mean an
orthonormal tetrad $e^i_{~(a)}(P) = \delta_a^i$ and its dual $e_k^{~(a)}(P) = \delta_k^a $
paralelly propagated along geodesics through $P$ ($P$ is the origin
of the {\bf NRC(P)}).
We have
\begin{equation}
e^i_{~(a)}(y) e_i^{~(b)}(y) = \delta_a^b.
\end{equation}
For a sufficiently small 4-dimensional domain $\Omega$ which
surrounds {\bf P} we require
\begin{equation}
\int\limits_{\Omega}{y^i d\Omega} = 0, ~~\int\limits_{\Omega}{y^i
y^k d\Omega} = \delta^{ik} M,
\end{equation}
where
\begin{equation}
M = \int\limits_{\Omega}{(y^0)^2 d\Omega} =
\int\limits_{\Omega}{(y^1)^2 d\Omega} =
\int\limits_{\Omega}{(y^2)^2
d\Omega}=\int\limits_{\Omega}{(y^3)^2 d\Omega},
\end{equation}
is a common value of the moments of inertia of the domain $\Omega$
with respect to the subspaces $y^i = 0,~~(i= 0,1,2,3)$.
We can take as $\Omega$, e.g., a  sufficiently small analytic ball centered
at $P$:
\begin{equation}
(y^0)^2 + (y^1)^2 + (y^2)^2 + (y^3)^2 \leq R^2,
\end{equation}
which for an auxiliary positive-definite metric
\begin{equation}
h^{ik} := 2 v^i v^k - g^{ik},
\end{equation}
can be written in the form
\begin{equation}
h_{ik}y^i y^k \leq R^2.
\end{equation}
A fiducial observer {\bf O} is at rest at the beginning {\bf P}
of the used Riemann normal coordinates {\bf NRC(P)} and its four-
velocity is $v^i =\ast~ \delta^i_o.$ $=\ast$ means that an
equations is valid only in special coordinates.
$\sigma(P;y)$ denotes the two-point {\it world function}
introduced in past by J.L. Synge \cite{Synge}
\begin{equation}
\sigma(P;y) =\ast {1\over 2}\bigl(y^{o^2} - y^{1^2} - y^{2^2}
-y^{3^2}\bigr).
\end{equation}
The world function $\sigma(P;y)$ can be defined covariantly by the
{\it eikonal-like equation} \cite{Synge}
\begin{equation}
g^{ik} \sigma_{,i} \sigma_{,k} = 2\sigma,
~~\sigma_{,i} := \partial_i\sigma,
\end{equation}
together with
\begin{equation}
\sigma(P;P) = 0, ~~\partial_i\sigma(P;P) = 0.
\end{equation}
The ball $\Omega$ can also be given by the inequality
\begin{equation}
h^{ik}\sigma_{,i} \sigma_{,k} \leq R^2.
\end{equation}
Tetrad components and normal components are equal at {\bf P}, so,
we will write the components of any quantity attached to {\bf P}
without tetrad brackets, e.g., we will write $S_a^{~b}(P)$
instead of $S_{(a)}^{~~~(b)}(P)$ and so on.

If $T_i^{~k}(y)$ are the components of an energy-momentum tensor
of matter, then we get from (5)
\begin{equation}
_m S_a^{~b}(P;v^l) = \bigl(2{\hat v}^l {\hat v}^m - {\hat g}^{lm}\bigr) \nabla_l \nabla_m {}
{\hat T}_a^{~b} = {\hat h}^{lm}\nabla_l \nabla_m {}{\hat T}_a^{~b}.
\end{equation}
Hat over a quantity denotes its value at {\bf P}, and $\nabla$
means covariant derivative.
Tensor $_m S_a^{~b}(P;v^l)$ is {\it the canonical superenergy tensor for matter}.

For the gravitational field, substitution of the canonical
Einstein energy-momentum pseudotensor as $T_i^{~k}$ in (5) gives
\begin{equation}
_g S_a^{~b}(P;v^l) = {\hat h}^{lm} {\hat W}_a^{~b}{}_{lm},
\end{equation}
where
\begin{eqnarray}
{W_a^{~b}}{}_{lm}&=& {2\alpha\over 9}\bigl[B^b_{~alm} +
P^b_{~alm}\nonumber\cr
&-& {1\over 2}\delta^b_a R^{ijk}_{~~~m}\bigl(R_{ijkl} +
R_{ikjl}\bigr) + 2\delta_a^b{\beta}^2 E_{(l\vert g}{}E^g_{~\vert
m)}\nonumber\cr
&-& 3 {\beta}^2 E_{a(l\vert}{}E^b_{~\vert m)} + 2\beta
R^b_{~(a\vert g\vert l)}{}E^g_{~m}\bigr].
\end{eqnarray}
Here $\alpha = {c^4\over 16\pi G} = {1\over 2\beta}$, and
\begin{equation}
E_i^{~k} := T_i^{~k} - {1\over 2}\delta_i^k T
\end{equation}
is the modified energy-momentum tensor of matter \footnote{In
terms of $E_i^{~k}$ Einstein equations read $R_i^{~k} = \beta
E_i^{~k}$.}.
On the other hand
\begin{equation}
B^b_{~alm} := 2R^{bik}_{~~~(l\vert}{}R_{aik\vert m)}-{1\over
2}\delta_a^b{} R^{ijk}_{~~~l}{}R_{ijkm}
\end{equation}
are the components of the {\it Bel-Robinson tensor} ({\bf BRT}),
while
\begin{equation}
P^b_{~alm}:= 2R^{bik}_{~~~(l\vert}{}R_{aki\vert m)}-{1\over
2} \delta_a^b{}R^{jik}_{~~~l}{}R_{jkim}
\end{equation}
is the Bel-Robinson tensor with  ``transposed'' indices $(ik)$.
Tensor $_g S_a^{~b}(P;v^l)$ is the {\it canonical superenergy
tensor} for gravitational field $\bigl\{^i_{kl}\bigr\}$.
In vacuum $_g S_a^{~b}(P;v^l)$ takes the simpler form
\begin{equation}
_g S_a^{~b}(P;v^l) = {8\alpha\over 9} {\hat h}^{lm}\bigl({\hat
C}^{bik}_{~~~(l\vert}{}{\hat C}_{aik\vert m)} -{1\over
2}\delta_a^b {\hat C}^{i(kp)}_{~~~~~(l\vert}{}{\hat C}_{ikp\vert
m)}\bigr).
\end{equation}
Here $C^a_{~blm}$ denote components of the {\it Weyl tensor}.

Some remarks are in order:
\begin{enumerate}
\item In vacuum the quadratic form $_g S_a^{~b}{}v^a v_b$, where $v^av_a = 1$, is {\it
positive-definite} giving the gravitational {\it superenergy density} $\epsilon_g$
for a fiducial observer {\bf O} which is at rest at the beginning
{\bf P} of the {\bf NRC(P)}.
\item In general, the canonical superenergy tensors are uniquely
determined only along the world line of the observer {\bf O}. But
in special cases, e.g., in Schwarzschild spacetime or in Friedman
universes, when there exists a physically and geometrically
distinguished four-velocity $v^i(x)$, one can introduce in an
unique way the unambiguous fields $_g S_i^{~k}(x;v^l)$ and $_m
S_i^{~k}(x;v^l)$.
\item We have proposed in our previous papers to use the tensor $_g S_i^{~k}(P;v^l)$
as a substitute of the non-existing gravitational energy-momentum
tensor.
\item It can easily seen that the superenegy densities
$\epsilon_g := _g S_i^{~k}v^iv_k, ~~\epsilon_m := _m S_i^{~k}v^i v_k$
for an observer {\bf O} who has the four-velocity $v^i$ correspond
exactly to the {\it energy of acceleration} ${1\over 2}m {\vec a}{\vec a}$
which is fundamental in Appel's approach to classical mechanics
\cite{Appel}.
\end{enumerate}

In past we have used the canonical superenergy tensors $_g S_i^{~k}$
and $_m S_i^{~k}$ to local (and also, in some cases, to global)
analysis of well-known solutions to the Einstein equations like
Schwarzschild and Kerr solutions; Friedman and Goedel universes,
and Kasner and Bianchi I, II universes.
The obtained results were interesting (See \cite {Gar1}).

We have also studied the transformational rules for the canonical
superenergy tensors under conformal rescalling of the metric
$g_{ik}(x)$\cite{Gar1,Gar2}.

The idea of the superenergy tensors can be extended on angular
momentum also \cite{Gar1}.The obtained angular superenergy tensors
do not depend on a radius vector and they depend only on {\it
spinorial part} of the suitable gravitational angular momentum
pseudotensor.\footnote{We have used in our investigation the
Bergmann-Thomson expression on angular momentum in general relativity.}

Changing the constructive definition (5) to the form
\begin{equation}
<T_a^{~b}(P)> := \displaystyle\lim_{\varepsilon\to
0}{\int\limits_{\Omega}\biggl[T_{(a)}^{~~~(b)}(y) -
T_{(a)}^{~~~(b)}(P)\biggr]d\Omega\over\varepsilon
^2/2\int\limits_{\Omega}d\Omega},
\end{equation}
where $\varepsilon := {R\over L}>0$ (equivalently $R = \varepsilon
L$) is a real parameter and $L$ is a dimensional constant : $[L] =
m$, one obtains {\it the averaged relative energy-momentum
tensors}.
Namely, for matter one obtains
\begin{equation}
<_m T_a^{~b}(P;v^l)> = _m S_a^{~b}(P;v^l) {L^2\over 6},
\end{equation}
and for gravity one obtains
\begin{equation}
<_g t_a^{~b}(P;v^l)> = _g S_a^{~b}(P;v^l) {L^2\over 6}.
\end{equation}

The components of the averaged relative energy-momentum tensors
have correct dimensions but they depend on a dimensional parameter
$L$ which plays role of a fundamental length.

Of course, the fundamental length $L$ must be infinitesimally
small because its existence violates local Lorentz invariance.
In \cite{Gar4} we have proposed  a universal choose of the
parameter $L$. Namely, we have proposed $L = 100L_P =\approx 10^{-33} m
$. Here $L_P := \sqrt{{\hbar G\over c^3}}\approx 10^{-35}m$ is the
{\it Planck length}. Following specialists in loop quantum gravity ({\bf LQG})
our $L= 100 L_P$ is approximately the smallest length over which the classical model of
the spacetime is admissible.

As we can seen, the averaged energy-momentum tensors differ from
the canonical superenergy tensors only by the constant multiplicator ${L^2\over
6}$, where $L$ means some fundamental length. Thus, from the
mathematical point of view these two kinds of tensors are
equivalent. Physically they are not because their components have
different dimensionality. Moreover, the averaged energy-momentum
tensors depend on a fundamental length $L$. Owing to the last fact
it seems that the canonical superenergy tensors are {\it more
fundamental} than the canonical averaged relative energy-momentum
tensors.This is the main reason why we have used (and still use)
superenergy tensors in our papers. But one should emphasize
that the canonical averaged relative energy-momentum tensors have
an important superiority over the canonical superenergy tensors:
their components have proper dimensions of the energy-momentum
densities.
\section{Energy and momentum carrying by gravitational waves}
In order to prove that any real gravitational wave transfers energy-momentum we
have used in our former papers the canonical superenergy tensor
$_g S_i^{~k}(P;v^l)$ for gravitational field.

By a direct calculation one can easily check
that this tensor gives {\it positive-definite} superenergy density
$\epsilon_s := _g S_i^{~k}v^i{}v_k$ and a {\it non-vanishing} superenergy flux
$P^i :=\bigl(\delta^i_k - v^i{}v_k\bigr)_g S_l^{~k}(P;v^a)v^l$ for every known solution of
the vacuum Einstein equations which represents a real gravitational wave, i.e., a wave
with $R_{iklm}\not=0$. Here $v^i$ means the four-velocity of an observer which is studying
gravitational field and who is at rest in a {\bf NRC(P)}.
As examples we have considered in \cite{Gar3,Gar4} the following gravitational waves
\begin{enumerate}
\item Linearly polarized, plane gravitational wave in the
coordinates $(U,V,X,Y)$ in which the line element reads
\begin{equation}
ds^2 = 2(Y^2 -X^2){F(U)\over 2}dU^2 + 2 dUdV -dX^2 - dY^2,
\end{equation}
where $F=F(U)$ is an arbitrary function;
\item Plane-fronted gravitational wave with parallel rays (p-p
wave) having the following line element in the coordinates $(U,V,X,Y)$
\begin{equation}
ds^2 = 2H(X,Y,U) dU^2 + 2dUdV - dX^2 - dY^2,
\end{equation}
where
\begin{equation}
\triangle H := \bigl({\partial^2\over\partial X^2}+
{\partial^2\over\partial Y^2}\bigr)H =0.
\end{equation}
The vector tangent to the V-lines is null and covariantly
constant.

The p-p vawe is a generalization of the plane wave.
\item The Einstein-Rosen (cylindrical) gravitational wave which
has the following line element in cylindrical coordinates
$ x^0 = ct,~x^1 = \varrho, ~x^2 = \varphi, ~x^3 = z$
\begin{equation}
ds^2 = e^{2(\gamma - \Psi)}\bigl(c^2 dt^2 - d\varrho^2\bigr)
-\varrho^2  e^{-2\Psi} d\varphi^2 - e^{2\Psi}dz^2.
\end{equation}
The metric functions $\gamma(x^0, x^1)$, $\Psi(x^0,x^1)$ satisfy
the following system of partial differential equations
\begin{eqnarray}
\Psi_{,11}& +& {1\over\varrho}\Psi_{,1}- \Psi_{,00} =0,\nonumber\cr
\gamma_{,1}&=& \varrho\bigl[(\Psi_{,1})^2 +
(\Psi_{,0})^2\bigr],\nonumber\cr
\gamma_{,0}&=& 2\varrho \Psi_{,0}{}\Psi_{,1}.
\end{eqnarray}
\end{enumerate}
For the all above gravitational waves we have obtained the
positive definite superenergy densities and non-null superenergy
fluxes (See \cite{Gar3,Gar4} for details).

The analogical result one gets also for any other real gravitational wave
in full agreement with the remark 1 of the previous Section.

It results from this that the every gravitational wave, which
has $R_{iklm}\not= 0$ {\it must also carry} the gravitational
energy-momentum. If not, then there would be a contradiction
between  an ``energy-momentum level'' and a ``superenergy level'',
because our canonical, gravitational superenergy tensor originated as a kind of
averaging in {\bf NRC(P)}of the canonical gravitational energy-momentum pseudotensor.

The following quasilocal constructions  confirm the above
statements.
Let us consider an observer {\bf O}  which is studying
gravitational field. His world-line is $x^a = x^a(s)$ and ${\vec
v}$: $v^a = {dx^a\over ds}$ represents his four-velocity. At any
point {\bf P}  of the world line one can  define an instantaneous,
local 3-space of the observer {\bf O} orthogonal to ${\vec v}$.
This instantaneous 3-space has the following interior proper
Riemannian metric
\begin{equation}
\gamma_{ab}:= v_a v_b - g_{ab}=\star  \bigl({g_{oa} g_{ob}\over
g_{oo}} - g_{ab}\bigr)= \star \gamma_{\alpha\beta} =\star (-)
g_{\alpha\beta},
\end{equation}
where the Greek indices run over the values $1,2,3$ (see e.g.,
\cite{LL}).

Then, by using the gravitational superenergy density $\epsilon_s$
and its flux $P^i$, one can easily construct in such instantaneous
local 3-space the following expressions which have proper
dimensions of the energy density and its flux
\begin{equation}
\epsilon_{en} := \oint\limits_{S_2} {\epsilon_s(P)d^2
S} \approx \epsilon_s(P) \oint\limits_{S_2}{d^2S} = 4\pi R^2\epsilon_s(P)>0,
\end{equation}

\begin{equation}
P^i :=\oint\limits_{S_2} {P^i(P) d^2S} \approx P^i(P) \oint\limits_{S_2}{d^2 S} =
4\pi R^2 P^i(P)\not = 0.
\end{equation}
Here $S_2$ means an infinitesimal sphere $\gamma_{\alpha\beta}x^{\alpha} x^{\beta} = R^2$
in the instantaneous local 3-space of the observer {\bf O} centered on this observer.

The expressions (30)-(31) give us the {\it relative gravitational
energy density} and its flux for an observer {\bf O} in his
instantaneous 3-space orthogonal to ${\vec v}$.

In our latter articles we have used with the same goal as above
the averaged relative gravitational energy-momentum tensor $<_g
t_a^{~b}(P;v^l)>$.

Namely, we defined

\begin{equation}
\epsilon_{en} := <_g t_a^{~b}(P;v^l)  v^a v_b,
\end{equation}
\begin{equation}
P^ i := \bigl(\delta^i_k - v^i v_k\bigr) <_g t_l ^{~k} (P;v^t)>
v^l.
\end{equation}

Here $v^i$ mean, as usual, the 4-velocity components of an
observer {\bf O} which is studying gravitational field.

From  the fundamental properties of the gravitational superenergy
and from the formulae (23)-(24) it is easily seen that
\begin{equation}
\epsilon_{en} >0, ~~P^i \not =0
\end{equation}
for every real gravitational wave.

Now, it seems to us that the using  of the canonical relative
gravitational energy-momentum tensor $<_g t_a^{~b}(P;v^l)> $ in
our considerations is more convincing than the using of the
canonical gravitational superenergy tensor $_g S_a^{~b}(P;v^l)$.
\section{Conclusion}
If one wants to get correct information about energy-momentum
(and angular momentum) of the real gravitational field by
application of coordinate-dependent pseudotensors and complexes,
then one has to use these strange objects in very special
situations and coordinates. For example, one can use these objects to global analysis
 of a closed system in asymptotically flat, {\it
Bondi-Sachs coordinates}
\cite{Moll}. In general one must use these objects locally in
Riemann normal coordinates {\bf NRC(P)} and extract from them
covariant, coordinate-independent information about gravitational field.
Our canonical gravitational superenegy tensor and our canonical averaged relative
energy-momentum tensor are exactly the quantities of such a kind. In application to
gravitational radiation these quasilocal quantities unambiguously
show that any real gravitational wave always transfer the energy
and momentum.

Thus, the negative conclusions given  by the authors
which have used pseudotensors and complexes in an arbitrary coordinates {\it are
incorrect}\cite{Gar3,Gar4}.

Finally, we would like to emphasize that the our local or, at
most, quasilocal results are complementary to the very old global
results obtained by A. Trautman \cite{Traut} in an asymptotically
flat spacetime which admitted outgoing gravitational radiation.

\acknowledgments
This paper was mainly supported by Institute of Mathematics
University of Szczecin.

\end{document}